\documentstyle[11pt]{article}
\titlepage
\newcommand{\p}{\underline}
\newcommand{\fr}{\frac}

\newcommand{\pr}{\partial}

\newcommand{\be}{\begin{equation}}

\newcommand{\ee}{\end{equation}}

\newcommand{\cH}{{\cal H}}
\newcommand{\bea}{\begin{eqnarray}}
\newcommand{\eea}{\end{eqnarray}}

\begin{document}


\title
{Description of Friedmann Observables in Quantum Universe }

\author{A.M.Khvedelidze\thanks{
Permanent address: Tbilisi Mathematical Institute,
Tbilisi, 380093, Georgia.} , V.V. Papoyan\thanks{
Permanent address: Yerevan State University, Yerevan, 375049, Armenia.} ,\\
Yu.G.Palii,
V.N.Pervushin\\[0.3cm]
{\normalsize\it Joint Institute for Nuclear Research},\\
 {\normalsize\it 141980, Dubna, Russia.}}

\date{\empty}

\maketitle
\medskip


\begin{abstract}
{\large
{The solution of the problem of describing the Friedmann observables
(the Hubble law, the red shift, etc.) in quantum cosmology is proposed
on the basis of the method of gaugeless Hamiltonian reduction in
which the gravitational part of the energy constraint is considered as a
new momentum. We show that the conjugate variable corresponding to the
new momentum plays a role of the invariant time parameter of evolution of
dynamical variables in the sector of the Dirac observables of the general
Hamiltonian approach. Relations between these Dirac observables and the
Friedmann observables of the expanding Universe are established for the
standard Friedmann cosmological model with dust and
radiation. The presented reduction removes an infinite factor from the
functional integral, provides the normalizability of the
wave function of the Universe and distinguishes the conformal frame of
reference where the Hubble law is caused by the alteration of the conformal
dust mass.}}

\end{abstract}

\vspace{2 cm}

\newpage

\section {\p { Introduction.}}

Hope  of solving fundamental problems of cosmology of the early
Universe by help of quantum gravity ~\cite{Dir58,ADM,Wheel,DeWitt,Faddeev}
has stimulated the development of the Hamiltonian approach to the theory of
gravity and cosmological models of the Universe.
A lot of papers  and some monographs (see e.g. \cite{Ryan1,Ryan2}) have
been devoted to the Hamiltonian description of cosmological models
of the Universe. The main peculiarity of the Hamiltonian
theory of gravity is the presence of nonphysical variables and constraints
due to the  diffeomorphism invariance of the theory.
Just this peculiarity is an obstacle for the
solution of the important conceptual  problems\\
-  interpretation of the wave function  and its non-normalizability ,\\
-  relations between the observational cosmology (the Hubble law and red
   shift) and the Dirac observables in  the Hamiltonian description of the
    classical and  quantum cosmologies.

One of the possible solutions of these problems  in the
Hamiltonian approach is to reduce the initial constraint system to the
unconstrained one by the full separation of pure gauge degrees of freedom from
physical ones ~\cite{Dirac}.
In the present paper, we would like to apply the
recently developed method of the Hamiltonian reduction of singular systems
 with the full separation of the gauge sector ~\cite{PhRevD,JMPh}
to a standard cosmological model of the Universe filled in by dust and
radiation to investigate the problems listed above and to compare
our reduced quantization with the extended approach
~\cite{Wheel,DeWitt,Faddeev}.

In section 2, the Hamiltonian version of standard model identical to classical
cosmology is given together with the Wheeler-DeWitt (WDW) equation which
follows from the model presented.

In section 3, we apply the Hamiltonian gaugeless reduction developed in
~\cite{PhRevD,JMPh} to construct the Dirac observables in the classical
theory.

Section 4 is devoted to quantization of the reduced system.

\section{\p {The Hamiltonian version of the Standard Model.}}

Let us consider the Friedmann standard model beginning from the
Hilbert -- Einstein action
\be\label{WEH}
W=\int\limits_{}^{}d^4x\sqrt{-g}\left[-\frac{{}^{(4)}R(g_{\mu\nu})}{16\pi G}
+{\cal L}_{matter}\right].
\ee
Following Friedmann we suppose the homogeneous distribution of the matter
described by the Lagrangian ${\cal L}_{matter}$ and, therefore, use the
 Friedmann -- Robertson -- Walker (FRW) metric
\be \label{FRW}  
(ds)^2=a^2(t)[N_c^2dt^2-\gamma_{ij}dx^idx^j].
\ee
Here $a(t)$ is a cosmic scale factor,
 $\gamma_{ij}dx^idx^j$ is the metric
of the three-dimensional  space of the constant curvature (we shall call it
the "conformal" one)
\be
{}^{(3)}R(\gamma_{ij})=\frac{-6k}{r_o^2} \;;\;\;\;k=0,\pm 1.
\ee
In this paper, we restrict ourselves to a closed space $k=+1$ to avoid
difficulties connected with an infinite volume and boundary conditions.
In this case, the parameter $r_o$ characterizes the volume of the
three-dimensional conformal space
\be
V_c=\int d^3x\sqrt{-\gamma}=2\pi^2r_o^3.
\ee
We kept  the variable $N_c$, in (\ref{FRW}),
in contrast with the Friedmann formulation of the standard model, where
the component $N_c$ is removed by the definition of the Friedmann time in
comoving frame
\be \label{tF}
dT=a(t)N_cdt.
\ee
Variable $N_c$ allows us to preserve the main peculiarity of the Einstein theory
(in the considered case)
-- the invariance with respect to  reparametrization of the coordinate time
\be \label{tt}
t\;\;\ss\;\;t'=t'(t)
\ee
and to reproduce the Einstein -- Friedmann equation for the homogeneous
distribution of dust and radiation by the variation of
the Einstein -- Hilbert action (\ref{WEH})  for the FRW metric
 ~\cite{Khved,PhLet}
\be\label{wham}
W=\int\limits_{}^{}dt
\left\{\sum_{l}P_l\dot A_l-\left[p_a\dot a-\frac{1}{2}\frac{d}{dt}
(p_aa)\right]-N_c\cH_{c}(p_a,a,\cH_R,M_D)\right\}.
\ee

The Lagrangian of the matter and
the corresponding energy $\cH_{c}$ are chosen to reproduce
the equation of state of radiation with dynamical variables $(P_l(t),A_l(t))$
and dust at rest (in the comoving frame (\ref{tF})) with the total mass $M_D$:
\be\label{hec}
\cH_{c}(p_a,a,\cH_R,M_D)=-\left(\frac{p_a^2}{2\beta}+
\frac{ka^2}{2r_o^2}\beta\right)+a(t)M_D+\cH_R
\ee
\be
\cH_R=\frac{1}{2}\sum_{l}^{}(P_l^2+\omega_l^2A_l^2).
\ee
The constant $\beta$ is
\be
\beta=V_{c}\frac{6}{8\pi G}=\frac{3\pi r_o^3}{2M_{Pl}^2}, \qquad
(M_{Pl}=1.22\;10^{19}GeV)
\ee

We kept here also the time surface term of action (\ref{WEH}).

It is easy to verify that the variation principles applied to the action
(\ref{wham}) reproduce the Friedmann evolution of the Universe
(filled in by matter with equation of state for dust and radiation)
in the comoving frame (\ref{tF}).

The equations of motion for variables of radiation $A_l,P_l$
\be
\dot A_l=N_c\{\cH_{c},A_l\}=N_c\frac{\pr\cH_{R}}{\pr P_l}\;\;;
\dot P_l=N_c\{\cH_{c},P_l\}=-N_c\frac{\pr\cH_{R}}{\pr A_l}
\ee
lead to the integral of motion
\be
\dot \cH_R=\frac{d}{dt}\cH_R=0\;\;\Rightarrow\;\;
\cH_R\left.\right|_{eq.m.}=E_R.
\ee
$E_R$ being a value of $\cH_R$ on the equations of motion.
The equation for variable $N_c$

\be \label{WNc}
\frac{\delta W}{\delta N_c}=0\;\;\Rightarrow\;\;\cH_{c}(p_a,a,E_R,M_D)=0
\ee
is treated as a constraint
and allows us to express the momentum $p_a$ in terms of the
cosmic scale factor
$a$ and parameters $E_R,M_D$
\be
p_{a(\pm)}=\pm\bar p(a,E_R,M_D)=\pm(2\beta)^{1/2}
[aM_D+E_R-\frac{a^2}{2r_o^2}]^{1/2}.
\ee
The evolution of the scale factor $a$ in the comoving frame (\ref{tF})
follows from the equation for $p_a$

\be \label{TFR}
\frac{\delta W}{\delta p_a}=0\;\;\;\Rightarrow\;\;
\frac{ada}{aNdt}\equiv\frac{ada}{dT}=p_a/\beta \;\;\Rightarrow\;\;
dT(E_R,a)=\frac{\beta ada}{\bar p(a,E_R,M_D)}
\ee
and completely reproduces the evolution of observables in the standard
Friedmann model of the Universe.
These observables are

the red shift as a function of the present time $T_o$
\be \label{zf}
z_0(T_o,d_F)=\frac{a(T_o)}{a(T_o-d_F/c)}-1=H_od_F/c+...
\ee
and a distance between the Earth and a cosmic object
\be\label{dF}
d_F(T)=a(T)d_c,
\ee
where $d_c$ is a distance in the conformal space with stationary metric
$\gamma_{ij}$ and
\be\label{H0}
H_o=\frac{1}{a(T_o)}\frac{d a(T_o)}{d T_o}
\ee
is the Hubble parameter.

On the other hand, the quantization of the scale factor
\be
i[p_a,a]=\hbar\;;\;\;p_a\ss\hat p_a=\frac{\hbar}{i}\frac{d}{da}
\ee
converts the energy balance equation (~\ref{WNc}) into the Wheeler -- DeWitt
equation ~\cite{Wheel,DeWitt}
\be\label{WDW}
[-\fr{\hbar^2}{2a\beta}\fr{d^2}{da^2}+\beta\fr{a}{2r_o^2}-
\fr{E_R}{a}-M_D]\Psi_{WDW}=0
\ee
(for the Einstein energy (\ref{hec}) in the comoving frame version
$\cH_E=\cH_{c}/a)$).

The main problem of our paper is to find the connection between the classical
Friedmann observables  (~\ref{tF}), (~\ref{zf}), (~\ref{H0}) and the wave
function of the Universe (~\ref{WDW}) and to establish a bridge between the
observational and quantum cosmologies.

\section{\p {The Dirac observables in classical and quantum theories.}}

The Einstein -- Hilbert action (~\ref{WEH}) (and, of course, its Hamiltonian
version (~\ref{wham}))
describes the first class constrained system according to the Dirac
classification ~\cite{Dirac}.

To construct the Dirac observables ~\cite{Dirac} of the first class
constrained systems, we fulfil gaugeless Hamiltonian
reduction ~\cite{PhRevD,JMPh} by using the canonical transformation to new
variables, so that the constraints become new momenta.

Thus, instead of the extended phase space $N_c,a, p_{a},A_l,P_l$ and the
initial action invariant under reparametrizations of the coordinate time
$(t \longmapsto t'=t'(t))$, we hope to get the reduced phase space which
contains only the fields of matter described by the reduced Hamiltonian.
These quantities (the fields of matter and the reduced Hamiltonian) are
invariant under the time reparametrizations and, therefore, are
the Dirac observables \cite{Dirac} by definition.

By means of the canonical transformation
\be\label{puas}
(p_a,a)\;\;\ss\;\;(\Pi,\eta)\;;\;\;\;\;\;\;
\{p_a,a\}|_{(\Pi,\eta)}=1
\ee
we convert the gravitational part of the Einstein -- Friedmann constraint
(~\ref{WNc}) to the new momentum
\be\label{pi}
\left(\frac{p_a^2}{2\beta}+\frac{ka^2}{2r_o^2}\beta\right)-aM_D=\Pi.
\ee
Equations (~\ref{puas}) and (~\ref{pi}) have two solutions (for $k=+1$)
\bea\label{a}
a(\Pi,\eta)&=&\pm\bar a(\Pi,\eta)\;;\;\;\;
\bar a(\Pi,\eta)=\left[\sqrt{2\Pi\frac{r_o^2}{\beta}}S(\eta)
+M_D\frac{r_o^2}{\beta}(1-C(\eta))\right],\\
p_a(\Pi,\eta)&=&\pm\bar p_a(\Pi,\eta)\;;\;\;\;
\bar p_a(\Pi,\eta)=
\left[\sqrt{2\Pi\beta}C(\eta)
+M_Dr_oS(\eta))\right], \label{pa}
\eea
where
\be
S(\eta)=\sin\frac{\eta}{r_o}\;;\;
C(\eta)=\cos\frac{\eta}{r_o}.
\ee
In terms of the new canonical variables $(\Pi,\eta)$ the action (~\ref{wham})
 for two solutions of eqs.(~\ref{puas}), (~\ref{pi}) has the forms
\be\label{wpm}
W^{(\pm)}=\int\limits_{}^{}dt
\left\{\sum_{l}P_l\dot A_l-
N_c(\cH_{R}-\Pi)\pm(\Pi\dot\eta+ \frac{M_D}{2} \frac{d}{dt}T(\Pi,\eta)) \right\},
\ee
where the function $T(\Pi,\eta)$,in the surface term
\be\label{TF}
T(\Pi,\eta)=\int\limits_{o}^{\eta}dx\bar a(\Pi,x).
\ee
coincides with the Friedmann time
 (~\ref{TFR}) in the parametric form (~\ref{a}),(~\ref{TF}), where $E_R$
is changed by $\Pi$.
So, we get the following equations of motion for the "cosmic sector"
$(\eta,\Pi,N_c)$
\be\label{np1}
\frac{\delta W^{(\pm)}}{\delta \eta}=0\;\;\;\Rightarrow\;\;\;\;\;
\dot\Pi=0
\ee
\be\label{np2}
\frac{\delta W^{(\pm)}}{\delta N_c}=0\;\;\;\Rightarrow\;\;\;\;\;
\Pi=\cH_R
\ee

\be\label{np3}
\frac{\delta W^{(\pm)}}{\delta \Pi}=0\;\;\;\Rightarrow\;\;\;\;\;
N_cdt=d\eta
\ee
and for "matter sector" ($A_l,P_l$).
\be\label{Al}
\frac{\delta W^{(\pm)}}{\delta B_l}=0\;\;\;\Rightarrow\;\;\;\;\;
\frac{d}{N_cdt} B_l=\{\cH_{R},B_l\}\;;\;\;\;\;\;\;(B_l=A_l,P_l)
\ee
Equations (~\ref{np2}), (~\ref{np3}) and (~\ref{Al}) mean that the new cosmological
variable $\eta$ converts into the invariant time parameter of evolution
of matter fields, and its canonical momentum $\Pi$ converts into the reduced
Hamiltonian. Equations (~\ref{np3}) determines the conformal time $\eta$.

The evolution of matter fields
can be be described by the reduced action (~\ref{wpm}) (on the solution of
equations (~\ref{np2}), (~\ref{np3}) of the cosmic sector)
\be\label{wred}
W^{red(\pm)}(\eta)=\int\limits_{o}^{\eta(t)}d\eta
\left\{\sum_{l}P_l\frac{dA_l}{d\eta}\pm\cH_{R}\right\}
\pm\frac{M_D}{2}T(\cH_R,\eta).
\ee
The last term follows from the surface term, and it does not influence
the equations of motion (~\ref{Al}).

The Dirac observables, here, are the "matter sector" (~\ref{Al}) and the
conformal time (~\ref{np3}). The partial variation of the action
(~\ref{wred}) (or (~\ref{wpm})) with respect to this time
(~\ref{np3}) represents the Tolman version ~\cite{Tolman} of the total energy
of the Universe filled in by radiation and dust in the conformal frame
of reference
\be\label{ED}
E_D^{(\pm)}=-\frac{\pr W^{(\pm)}}{\pr\eta(t)}=
\pm(E_R+\frac{M_D}{2}a(E_R,\eta)).
\ee
In this frame interval is determined as
\be \label{DSD}  
(ds)^2_D=d\eta^2-\gamma_{ij}dx^idx^j\;;\;\;\;\;
\ee
with a stationary space distance, in contrast with the comoving frame
(~\ref{FRW}) with the Friedmann time (~\ref{tF}) and observable interval
\be \label{DSF}  
(ds)^2_F=dT^2-a^2(T)\gamma_{ij}dx^idx^j.
\ee

The variation of the reduced action (\ref{wred}) with respect to the
Friedmann time (\ref{TFR}) leads to the Tolman version ~\cite{Tolman}
of the Friedmann energy
\be\label{EF}
E_F^{(\pm)}=-\frac{\pr W^{(\pm)}}{\pr T(\eta)}=
\pm\left(\frac{E_R}{a(E_R,\eta)}+\frac{M_D}{2}\right).
\ee
The first term of this energy describes the conventional Friedmann evolution
of the red shift of "photons" (\ref{zf}) in the process of expansion of the
Universe in the comoving frame (\ref{DSF}).

Thus, the considered version of the Hamiltonian reduction also
describes the conventional observables of the classical
FRW cosmology provided the choice of the comoving frame (\ref{DSF})
(this means the choice of a corresponding observer).
The Friedmann observer, in the comoving frame, sees stationary mass, while
the wavelengh of a photon (therefore, the energy of radiation) are changing
according to the red shift law (\ref{zf}).

The Dirac observer, in the conformal frame (\ref{DSD}), sees
constant wavelengh of a
photon, while all masses in the Universe are changing so that the spectrum of an
atom on a cosmic object at moment of emission $\eta_o-d_D/c $ differs from the
Earth spectrum at moment of observation $\eta_o$.

As it has first been  established by Hoyle and Narlikar ~\cite{Narlik} the
red shift also exists in terms of the conformal time
\be \label{zD}
z_o=\frac{ma(\eta_o-d_D/c)}{ma(\eta_o)}-1=
\frac{a(\eta_o-d_D/c)}{a(\eta_o)}-1
\ee
We see that in any frame of reference the observables are "energy" and
"time", but not the cosmological scale factor $a$ and its momentum $p_a$.
This fact explains the difficulty in comparing the Friedmann
observables "energy - time" with the result of the
WDW quantization in terms of $a,p_a$.

\section {\p { Quantization.}}

Let us quantize the theory (~\ref{wpm}). First at all note that the
commutation relation $i[\hat \Pi,\eta]=\hbar$ and the WDW equation in terms
of the new variables
\be\label{PS}
(\hat \Pi-\cH_R)\Psi_{WDW}=0
\ee
do not reflect all information of the classical theory, in particular, about
two different solutions (\ref{wpm}) with signs $(\pm)$ and the dust evolution
of the "observable red shift" (\ref{zD}). The latter is hidden in the
surface term which contributes to the total energy (\ref{ED}).

These peculiarities of the classical theory can be described, if we use
the action (\ref{wpm}) to determine the momentum $\Pi$.
\be\label{Pi}
\Pi^{(\pm)}=\frac{\pr L^{(\pm)}}{\pr\dot\eta}=\mp(\Pi+\frac{M_D}{2}
\bar a(\Pi,\eta)).
\ee
We should use the class of the wave functions where the constraint (\ref{PS})
is fulfilled.

Finally, using the operator
$$
\hat\Pi^{(\pm)}=\pm\frac{\hbar}{i}\frac{d}{d\eta}
$$
we get for the wave function the following spectral decomposition
\be\label{PSD}
\Psi_D(\eta|A_l)=\sum_{E_R}^{}
\left[\alpha_{E_R}^+\Phi^{(+)}_{E_R}(\eta)<E_R|A_l>+
\alpha_{E_R}^-\Phi^{(-)}_{E_R}(\eta)<E_R|A_l>^*\right]
\ee
where $<E_R|A_l>$ are the production of the Hermite polynomials,
\be\label{PhE}
\Phi^{(\pm)}_{E_R}(\eta)=exp\{\pm i(E_R\eta+\frac{M_D}{2}\bar T(E_R,\eta)\},
\ee
and $\alpha^{(\pm)}$ are the operators of creation of the Universes with
the total energy (~\ref{ED})
\be\label{ER}
E_R=\sum_{l}^{}\omega_l\; n_l,
\ee
and $n_l$ is quantum number of occupation   of a "photon" with the energy
$\omega_l$.
We can express the observable red shift (\ref{zD}) in terms of the wave
function with quantum numbers (\ref{ER})
\be \label{zM}
z_o=\frac{{\cal M}_{E_R}(\eta_o)}{{\cal M}_{E_R}(\eta_o-\fr{d_D}{c})}-1
\ee
where
\be
{\cal M}_{E_R}=\left[\Phi^{(+)*}_{E_R}(\eta)\frac{\pr}{i\pr \eta}
\Phi^{(+)}_{E_R}(\eta)-E_R\right]
\ee

The variation of the wave function (\ref{PhE}) with respect to the Friedmann
time (\ref{TFR}) leads to the energy of the red shift of a "photon" in the
comoving frame of the Friedmann observer.

\be\label{EQPhi}
 \frac{d}{idT(\eta)} \Phi^{(\pm)}_{E_R}(\eta)=
\pm\left(\frac{E_R}{a(T)}+\frac{M_D}{2}\right)\Phi^{(\pm)}_{E_R}(\eta)                        .
\ee
In contrast with the WDW wave function (~\ref{WDW}), eq. (~\ref{EQPhi})
establishes the direct relation of the Dirac wave function to the
observables of the classical theory.

The wave function (~\ref{PSD}) is normalizable for variables of the Dirac
physical sector  $(A_l,P_l)$ and has clear physical interpretation.
The wave function of the Universe is the conventional wave function of
massless excitations at the conformal time multiplied by the
wave function of a particle at the Friedmann time with a half mass of
the Universe.

The corresponding functional integral representation of the Green function
does not contain functional integration over the variable $\eta$ (as it
was excluded from the Dirac sector of physical variables) ~\cite{gog}.
This conversion of the variable into the time parameter
excludes the infinite gauge factor from the functional integral
discussed by Hartle and Hawking ~\cite{hart}.


\section{ {Interpretation and conclusion.}}

The aim of the present paper is to investigate relations between the
Friedmann cosmological observables and the Dirac physical ones
in the Hamiltonian approach to quantization of the Universe using a simple
but important example of the homogeneous Universe filled in by dust and
radiation.

An essential difference of the research presented here from the analogous
papers on the Hamiltonian dynamics of cosmological models is complete
separation of the sector of physical invariant variables from the pure gauge
sector by application of the gaugeless reduction ~\cite{PhRevD,JMPh}.
The main point is that in the process of reduction
one of variables converts into the observable invariant time.
We have shown that this conversion of the variable to the time parameter
leads to the normalizability of the wave function of the Universe
and plays the role of gauge-fixing for removing an infinite factor from the
corresponding functional integral.
The considered reduction allows us to give the definite
mathematical and physical treatment of the wave function and clarifies
its relation to the observational cosmology.

The choice of the conformal frame of reference was a crucial to construct
the Dirac "observables" in the generalized Hamiltonian approach.
These "observable" are connected  with the
Friedmann observables by conformal transformations with the cosmic
scale factor. However, these transformations have singularity
at the beginning of the Dirac time.
From this point of view, the Dirac "observables" in the frame connected with
the radiation seems to be more
fundamental than the Friedmann "observables" in the comoving frame
connected with massive dust.

At the beginning of the Universe (at time $\eta\ss 0$) the energy of
particles at rest (forming the dust) in a closed space becomes larger than
their masses, and all dust converts into massless excitations with wavelenghs
of an order of a conformal size of the Universe $r_0$. For these excitations
the region of validity of quantum theory coincides with the size of the
Universe. Thus, at the beginning of the Universe,  the dust disappears and
the Friedmann observables connected with massive dust lose physical meaning.
While the Dirac "observer" sees
the closed space filled by the  homogeneous massless excitations bounded all
regions in the Universe. For the Dirac "observer"  the difficulties of
singularity and horison do not exist.

\vspace{0.5cm}
\newpage

Acknowledgments.

We are happy to acknowledge interesting and critical
discussions with Profs. Cecile DeWitt-Morette, Z.Perjes
and to thank the Russian Foundation
for Basic Investigation, Grant N 96\--01\--01223 for support.


\begin{thebibliography}{99}
\bibitem{Dir58} P.A.M.~Dirac. Proc.Roy.Soc., \underline {A~246} (1958) 333;
Phys.Rev. \underline{114} (1959) 924.
\bibitem{ADM} R.~Arnowitt, S.~Deser, C.W.~Misner. Phys.Rev.
\underline {117} (1960) 1595.
\bibitem{Wheel} J.A.Wheeler. In Batelle Recontres :
1967 Lectures  in Mathematics and Physics, edited by  C. DeWitt and
J.A.Wheeler, Benjamin, New York, (1968).
\bibitem{DeWitt} B.S.~DeWitt. Phys.Rev. \underline{160} (1967) 1113.
\bibitem{Faddeev} L.D.~Faddeev,V.N.~Popov. Usp. Fiz. Nauk 111 (1973) 427.
\bibitem{Ryan1} M.P.~Ryan, Jr., and L.C.~Shapley. "Homogeneous Relativistic
Cosmologies", Princeton Series on Physics, Princeton University
Press, Princeton, N.Y. 1975.
\bibitem{Ryan2}M.P.~Ryan, "Hamiltonian Cosmology",
Lecture Notes in Physics N 13 Springer Verlag,
Berlin--Heidelberg--New York, 1972.
\bibitem{Dirac} P.A.M.~Dirac, Lectures on Quantum Mechanics, Belfer Graduate
School of Science Yeshiva University, New York, 1964.
\bibitem{PhRevD} S.A. Gogilidze, A.M. Khvedelidze, V.N. Pervushin.
Phys. Rev.D 53 (1996) 2160.
\bibitem{JMPh} S.A. Gogilidze, A.M. Khvedelidze, V.N. Pervushin.
J.Math.Phys. 37 (1996) 1760.
\bibitem{Khved} A.M.Khvedelidze, V.V.Papoyan, V.N.Pervushin. Phys.Rev.D
{\p \bf 51}, (1995) 5654.
\bibitem{PhLet} V.Pervushin, V.Papoyan, S.Gogilidze, A.Khvedelidze, Yu.Palii,
V.Smirichinski,\\ Phys.Lett.{\bf B365,}(1996), 35
\bibitem{Tolman} R.C.~Tolman,{ \it Relativity, Thermodynamics and Gravitation }
(Calderon Press, Oxford, 1969);\\
R.C.~Tolman, Phys.Rev. {\bf 35}, 875 (1930).
\bibitem{Narlik} J.V.~Narlikar in "Astrofizica e  Cosmologia, Gravitazione,
Quanti e Relativita", G. Barbera, Firenze , 1979.
\bibitem{gog}
S.A. Gogilidze, A.M. Khvedelidze, V.V. Papoyan, Yu.G. Palii,
V.N. Pervushin,\\
``Dirac and Friedmann Observables in Quantum Universe with Radiation.''
{\it Preprint JINR, E2-96-475, Dubna, 1996,
submitted to "Gravtation and Cosmology".}
\bibitem{hart}
J.B. Hartle, S.W. Hawking, Phys. Rev. D {\p \bf 28} (1983), 2960.
\end{thebibliography}
\end{document}